\documentclass[12pt]{article}
\usepackage{amsmath}
\usepackage{amssymb}
\usepackage{graphicx}
\usepackage{amsfonts}
\usepackage{color}
\usepackage{soul}
\usepackage{wrapfig}
\usepackage{titlecaps}
\usepackage{ragged2e}
\usepackage{pifont}
\usepackage{nccmath}
\usepackage{graphicx}
\graphicspath{ C:\Users\jdubi\Documents\PhD\Project ManyBodySystems\Paper\Draft1}
\usepackage{epstopdf}
\usepackage{float}
\usepackage{wrapfig}
\usepackage{setspace}
\usepackage{geometry}
\usepackage{braket}
\usepackage{lineno}
\usepackage{xfrac}

\makeatletter
\def\captionof#1#2{{\def\@captype{#1}#2}}
\makeatother
\makeatletter
\newcommand*{\addFileDependency}[1]{% argument=file name and extension
  \typeout{(#1)}
  \@addtofilelist{#1}
  \IfFileExists{#1}{}{\typeout{No file #1.}}
}
\makeatother

\def\beq{\begin{equation}}
\def\eeq{\end{equation}}
\def\beqa{\begin{eqnarray}}
\def\eeqa{\end{eqnarray}}

\newcommand{\appropto}{\mathrel{\vcenter{
			\offinterlineskip\halign{\hfil$##$\cr
				\propto\cr\noalign{\kern2pt}\sim\cr\noalign{\kern-2pt}}}}}

\usepackage{scicite}

\usepackage{times}
\newenvironment{sciabstract}{%
\begin{quote} \bf}
{\end{quote}}

\title{Do Photosynthetic  Complexes Use Quantum Coherence to Increase Their Efficiency? \\ Probably Not}

\author
{Elinor Zerah Haursh,$^{1,2}$ Yonatan Dubi,$^{\ast 1,2}$\\
\\
\normalsize{$^{1}$Department of Chemistry and $^2$Ilse-Katz Institute for Nanoscale Science and Technology,}\\
\normalsize{ Ben-Gurion University of the Negev, Beer-Sheva 84105, Israel}\\
\\
\normalsize{$^\ast$To whom correspondence should be addressed; E-mail:  jdubi@bgu.ac.il.}
}

\date{}

\begin{document} 

% Double-space the manuscript.

\baselineskip24pt

% Make the title.

\maketitle

% Place your abstract within the special {sciabstract} environment.
\begin{sciabstract}
 {Answering the titular question has become a central motivation in the field of quantum biology, ever since the idea was raised following a series of experiments demonstrating wave-like behavior in photosynthetic complexes. Here, we report a direct evaluation of the effect of quantum coherence on the efficiency of three natural complexes. An open quantum systems approach allows us to simultaneously identify their level of "quantumness" and efficiency, under natural physiological conditions. We show that these systems reside in a mixed quantum-classical regime, characterized by dephasing-assisted transport. Yet, we find that the change in efficiency at this regime is minute at best, implying that the presence of quantum coherence does not play a significant role in enhancing efficiency. However, in this regime efficiency is independent of any structural parameters, suggesting that evolution may have driven natural complexes to their parameter regime in order to "design" their structure for other uses.}   
\end{sciabstract}

% In setting up this template for *Science* papers, we've used both
% the \section* command and the \paragraph* command for topical
% divisions.  Which you use will of course depend on the type of paper
% you're writing.  Review Articles tend to have displayed headings, for
% which \section* is more appropriate; Research Articles, when they have
% formal topical divisions at all, tend to signal them with bold text
% that runs into the paragraph, for which \paragraph* is the right
% choice.  Either way, use the asterisk (*) modifier, as shown, to
% suppress numbering.

\section*{Introduction}
In the photosynthetic process, energy is transferred from an antenna (where light is collected) to a reaction center (where the energy is converted to chemical energy, to be used later by the organism). Excitons -bound electron–hole pairs - are the energy carriers in the photosynthetic process, carrying the harvested solar energy from the antenna to the reaction center, through a network of Bacteriochlorophylls (BChls), the so-called exciton-transfer complex (ETC) \cite{mohseni2014quantum}. Interest in the dynamics of excitons in the ETC exploded over the last decade, following recent experiments, where ultrafast nonlinear spectroscopy signals showed long-lived oscillations \cite{Lee2007,engel2007,Calhoun2009,Collini2010,Panitchayangkoon2010,Panitchayangkoon2011,blankenship2011comparing}. 
The discovery of coherent oscillations in ETCs pushed forward the hypothesis that in natural photosynthetic complexes, which are extremely efficient, quantum coherence in the presence of an environment is used to assist energy transfer, an idea that has generated much excitement (and debate) \cite{wu_efficient_2012,Leon-Montiel2014, rathbone2018coherent,Fleming2011,kassal2013,duan2017nature,chenu2015coherence,keren2018photosynthetic,cao2020quantum,bourne2019structure,yang_steady_state_2020}. \\
The problem can basically be summarized in two seemingly simple questions; (1) can quantum coherence exist during the biologic process of photosynthetic energy transfer? (2) if it does, is it used in some way by the natural system to enhance its efficiency? The latter question is actually more subtle, and perhaps better phrased as (see Fig. 1(a)): does the presence of quantum coherence adds any functional advantage, such that it played a role in the driving forces that led, through evolution, to the current design of the natural photosynthetic apparatus? \\
Many theoretical (and experimental) works have addressed these questions, yet the question in the title remains largely unanswered. One reason is that while experiments are performed in vitro with coherent (pulsed) light, natural systems operate under very different conditions, namely continuous incoherent excitation \cite{Manzano2013,Pelzer2014,pachon2017open,brumer2018shedding,morales2020photochemical,bourne2019structure}, and observing coherence under natural conditions is a very challenging task. That and more, it is hard to make the connection between the observed experimental findings and the energy transfer {\sl efficiency}, which is related to the total rate at which energy can flow from the antenna to the reaction center (two ingredients which are essentially absent in the experiments).\\  
Here we address the aforementioned questions, using tools developed from the theory of open quantum systems. Our approach allows us to evaluate efficiency directly while taking relevant physical parameters into account, and provides a simple way to estimate if the system is "quantum" or "classical", i.e. to evaluate whether environment-induced dephasing has pushed the system into the classical regime. We find that the answers to the questions posed above are "yes" and "no", namely that the excitonic system is indeed in the quantum-coherent regime (even for fast dephasing of $\sim 100$fs \cite{thyrhaug2018identification}), but that quantum coherence has only a minute effect on transport efficiency. Put simply, our findings suggest that the answer to the question posed in the title is negative. 
%\section{Model and Method}
%\section{Results}
\section*{ Methods} 
The approach we take enables us to calculate both the total exciton current through the ETC, and the exciton population at each BChl site simultaneously, under the condition of continuous incoherent excitaions \cite{brumer2018shedding,morales2020photochemical}, with physiologically relevant parameters. The joint evaluation of both currents and populations allows us to answer the questions posed in the introduction. First, the exciton current is a direct measure of the ETCs efficiency. This is easy to understand; the efficiency is simply the ratio between power output and power input. The power input is constant, and the power output is essentially the exciton energy times the exciton current. \\
Second, the populations allow us to evaluate the level of "quantumness" of the system. This was recognized in recent work \cite{zerah2018universal}, where a connection between exciton population, dephasing rate and the approach to classicality was established, through the mechanism of environment-assisted quantum transport (ENAQT). When an environment acts on a disordered quantum network (such as the exciton transfer complex), it induces a finite dephasing time. ENAQT is the situation where the dependence of quantum transport on the dephasing rate is non-monotonic, showing a maximum at some optimal dephasing rate, and was considered to be a possible mechanism for the high efficiency of photosynthetic complexes \cite{Caruso2009,Rebentrost2009,Mohseni2008,Caruso2009,Chin2010,Plenio2008,Kassal2012,Caruso2014,Li2015,Marais2013,Sinayskiy2012,Chen2015,dubiinterplay2015,Berman2015,Baghbanzadeh2016,cao2009optimization,sarovar2010}. 
The relation between ENAQT and particle populations is as follows %(Fig.~\ref{fig:schem}(b)) \cite{zerah2018universal}
. At the quantum regime (very small dephasing rate), the populations are essentially determined by the Hamiltonian structure of the network and the positions of the source and drain (antenna and reaction center in photosynthetic complexes). As dephasing rate increases, it reduces the variations in populations, so-called "population uniformization", flattening the population distribution, and resulting in an increase in current, which reaches a maximum at some optimal rate. For high dephasing rates, the system becomes essentially classical, and the populations are organized according to Fick's law, i.e. the formation of a uniform gradient between the source and drain. It is this gradient which can be used to define how "classical" is the system; once the gradient is fully formed (such that increasing the dephasing rate no longer changes the populations), one can say that the system is classical. The population uniformization mechanism is depicted in Fig. 1(b).

\begin{figure}[H]
\includegraphics[width=1\linewidth]{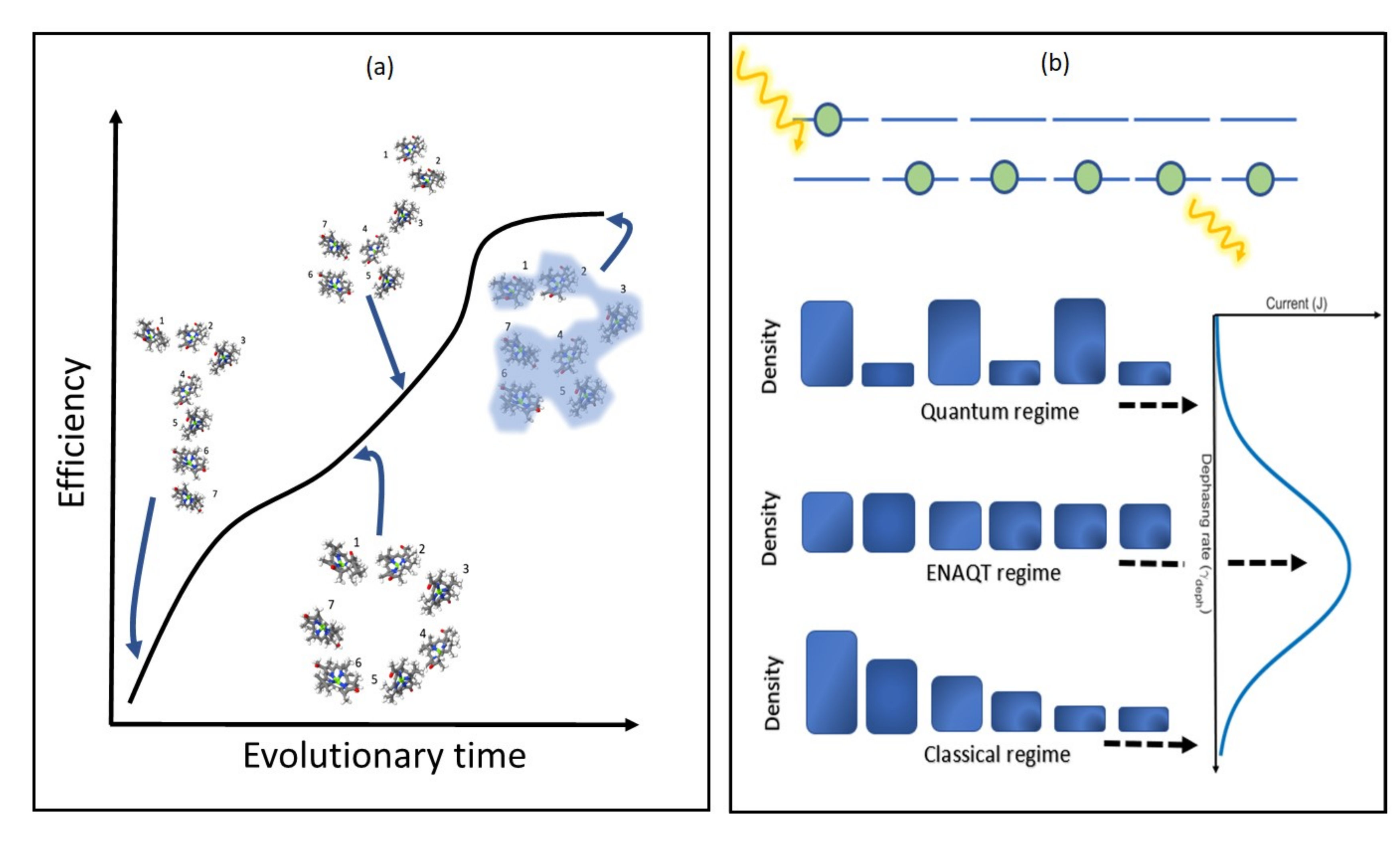}
\caption{Mechanisms of efficiency-driven evolution and environment-assisted quantum transport. (a) Schematic description of the evolutionary progress of photosynthetic complexes towards their current geometry, with efficiency being the evolutionary driving force. As evolution progresses, the structure of the photosynthetic complex evolves towards its current structure (the FMO complex in this example), while increasing efficiency. Is this indeed the evolutionary pathway of photosynthetic complexes, and if so is quantum coherence part of the efficiency enhancement is a central question in the field of quantum biology. 
(b) Schematic depiction of the population uniformization mechanism shown for a uniform chain of six sites (Blue lines depict the sites in the chain, Yellow arrows show the excitation of first site and extraction from fifth site). The density of the sites are described by blue bars for the quantum regime,  ENAQT regime and  classical regime, along with a schematic form for the current vs dephasing curves. 
 }\label{fig:schem}
\end{figure}

In what follows, we evaluate both the transfer efficiency and populations from the Linbdlad quantum master equation \cite{breuer2002theory}, taking physically-relevant parameters (detailed form of the Lindblad equation is given in the SM section I).
The dephasing rate serves as a free parameter in the calculations, but can be evaluated from experiments to be in the range of 10-$10^3$ femtoseconds. \cite{Lee2007,engel2007,Calhoun2009,Collini2010,Panitchayangkoon2010,Panitchayangkoon2011}. \\
We consider three different photosynthetic exciton transfer complexes; the Fenna-Matthews-Olson (FMO) complex which appears in Green sulfur bacteria, the PC-645 protein, which is a sub-unit of the photosynthetic apparatus in cryptophyte algae, %[www.pnas.org/content/111/26/E2666]
 and LH2, part of the photosynthetic apparatus of the purple photosynthetic
bacterium {\sl Rhodopseudomonas acidophila} (their schematic structures are plotted in the insets to Figs.2 (a), 2 (b) and 4, respectively). All three complexes were shown to exhibit coherent energy-transfer oscillations in non-linear 2D spectroscopy measurements  \cite{rathbone2018coherent,collini_coherently_2010,hildner_quantum_2013,ma2017excitonic,ferretti2014nature,dean2016vibronic,Panitchayangkoon2010,Lee2007}. The Hamiltonian parametrization of each complex was taken from previous literature \cite{Cho2005,hein_modelling_2012,tretiak_exciton_2000,tretiak_bacteriochlorophyll_2000,mirkovic_ultrafast_2007}, and are provided in the SM (section I).{ Some crystallographic measurements suggest an updated model of the FMO complex, containing eight bacteriochlorophylls (Bchls) instead of seven \cite{ben-shem_evolution_2004,tronrud_structural_2009}, a structure which has been parametrized and studied in the context of FMO energy transfer (e.g. \cite{ritschel2011absence,moix_efficient_2011,dubi_interplay_2015}). Here we chose to focus on the seven Bchls model, as previous works demonstrated that the expected difference between the two models would be insignificant \cite{dubi_interplay_2015,moix_efficient_2011}.   }\\
The remaining parameters which are needed to fully define the parameter set are injection and extraction rates, i.e. the rate at which excitons are pushed into the ETC and extracted to the reaction center. The extraction rate can be estimated by considering the time-scales of different transport processes that take place in the photosynthetic complexes. For instance,  the exciton transfer time between  adjacent LH2 complexes was found to be $3-100$ps \cite{rathbone2018coherent, hess1995temporally,engel2007,duan2017nature,mirkovic_ultrafast_2007,hildner_quantum_2013,bourne2019structure}, and the trapping time of energy by the core complexes in PC-645 was found to be $\approx 100$ps  \cite{van_der_weij-de_wit_phycocyanin_2008} We then set the extraction rate to an average of $\gamma_{ext}=0.1$ ps$^{-1}$. However, a range of extraction rates was tested, and our results and conclusions are essentially insensitive to the extraction rate, as long as it is much larger than the injection rate (see below).\\
The injection (or excitation) rate is limited by the absorption cross section, which was estimated for  by evaluating that there are $\sigma \sim 14$/s (14 excitons per second) for biological intensity of $I\sim 18\frac{W}{m^2}$ \cite{geyer2006reconstruction}. The sunlight intensity can be as high as $I_{max}\sim 1300\frac{W}{m^2}$ %[Stellmacher2000]
(on a bright day at the equator), which can be absorbed by $N\approx400$ complexes in one vesicle \cite{scheuring2004watching,scheuring2004variable}. The resulting {\sl upper limit} for the injection rate is then $\gamma_{inj}=N \times \sigma \times I_{max} / I\sim 0.4\mu $s$^{-1}$. { We note that we have used the same injection rate for all organisms here, although FMO probably does not absorb energy directly from the sun. However, we considered the maximum injection rate that can obtained from the baseplate. This ensures that our results are correct even for extreme conditions (i.e. give an upper bound).}
\section*{ Results}
\textbf{Currents and populations in FMO and PC-645}\\
With all parameters set, one can now evaluate the effect of the environment on the photosynthetic transfer efficiency. In Fig. 2 we plot the exciton current as a function of dephasing rate, for the FMO complex (a) and the PC-645 complex (b). Insets are the schematic structures of the complexes, respectively. The similarity between the plots is an indication for the relative insensitivity of the current to the internal structure Hamiltonian \cite{sener2002robustness}. The green-shaded area in Fig.2 shows the region of physiological dephasing rates.  The ENAQT effect is clearly visible, as the current shows a maximum in the dephasing rate. However, the enhancement in the current due to dephasing is minute, constituting only $\sim 0.0015$ percent increase (even taking extreme values for the injection and extraction rates yields an ENAQT enhancement of only a few percent, see SM Section III). It seems unlikely that such a small efficiency enhancement would be a meaningful evolutionary driving force; it is more likely that other factors were prominent in the evolutionary design of these photosynthetic complexes. \\
\begin{figure}[H]
\includegraphics[width=1\linewidth]{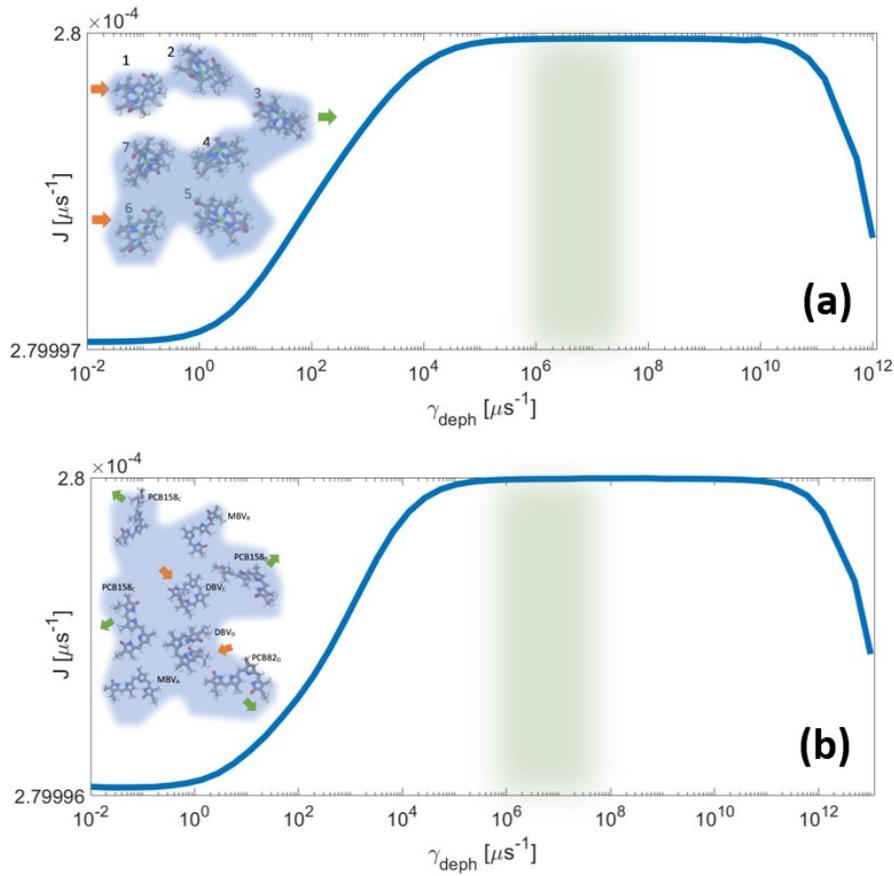}
\caption{
Calculated exciton current as a function of dephasing for the FMO (a) and pc-645 (b) complexes. The shaded Green area indicates the estimated range of physiological dephasing rates. Insets show a schematic description of the exciton complexes (the full Hamiltonians used are provided in the SM section VI). }
\label{fig:bioRng}
\end{figure}
The structure of the current-dephasing rate dependence already gives a hint that both FMO and PC-645 operate in the ENAQT regime under natural conditions. This can be further corroborated by looking at the exciton populations within the transfer complex for different dephasing rates. As pointed above, the three regimes (quantum, ENAQT and classical regimes) have very distinct features; the quantum regime is characterize by a spread of the populations determined by the structure of the Hamiltonian, the ENAQT regime by uniform populations, and the classical regime by a linear population gradient from source to drain. \\
In Fig. 3 the exciton population of the FMO complex is plotted for three values of dephasing rate, corresponding to the quantum ($\gamma_{deph}=10^{-4}\mu $s$^{-1}$), biological conditions  (ENAQT regime, $\gamma_{deph}=10^{6}\mu $s$^{-1}$) and classical regime ($\gamma_{deph}=10^{12}\mu $s$^{-1}$). Fig. 3(a) shows the occupation as a function of site number, but since the FMO is not a simple linear chain, in Fig. 3(b-d) we show the population on the FMO lattice, color-coded such that brighter colors represent lower density.\\
One can clearly observe the population uniformization that leads to ENAQT; In the quantum regime, populations seem disordered, and are determined by the interplay between the structure of the wave-functions and the source and drain positions. At intermediate dephasing, the population is essentially uniform, and a uniform gradient is formed between source and drain (sites 6 and 3) for strong dephasing.  \\

\begin{figure}[H]
\centering
\includegraphics[width=1\linewidth]{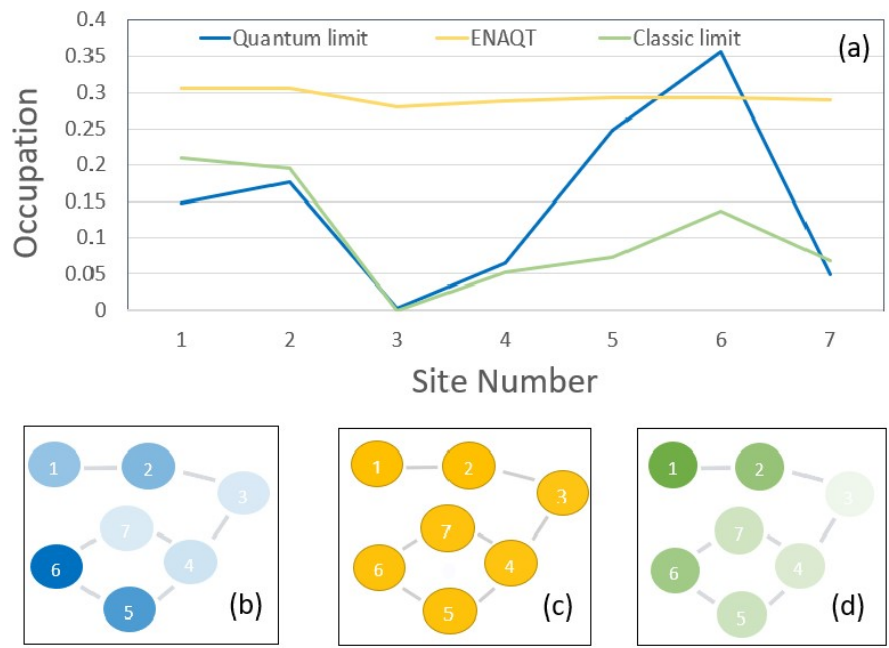}
\caption{Calculated exciton current as a function of dephasing for the FMO (a) and pc-645 (b) complexes. The shaded Green area indicates the estimated range of physiological dephasing rates. Insets show a schematic description of the exciton complexes (the full Hamiltonians used are provided in the SM section VI).}
\label{fig:FMO_Occup}
\end{figure}
\textbf{ENAQT in LH2}\\
In the LH2 complex, making the connection between ENAQT and population (i.e. recognizing the diffusive regime by observing a population gradient) is harder, because there is no simple spatial separation between the antenna (injection sites) and reaction center (source site) such that a gradient can be identified.   As depicted in the inset of Fig. 4, the LH2 complex is composed of two rings of BChl pigments, B800 (yellow ring) and B850 (blue ring), named after their energy absorption resonance  (in nm), connected by a ring of Lycopen molecules (gray, long molecules)  that absorb energy in the visible region of the spectrum \cite{mcdermott1995crystal,isaacs_light-harvesting_1995,koepke1996crystal,tretiak_exciton_2000}. Each of these parts can absorb light that excites an exciton that later would be transferred from one of the rings to the reaction center \cite{isaacs_light-harvesting_1995,hess1995temporally,mirkovic_ultrafast_2007}. 
The structure thus enables the occurrence of many exciton transfer paths. Nevertheless, a current vs dephasing curve for LH2 can still reveal the importance (or lack thereof) of coherence in transport.\\
In order to evaluate the efficiency of energy transfer in LH2 and its dependence on the dephasing rate, we calculate the excitonic current through LH2, considering multiple paths. Specifically, we assume that an exciton can be excited and extracted in any one of the BChl or molecular sites. In Fig. 4 we plot current as a function of dephasing rate for the LH2 system. Light pink lines are examples of specific paths, and the solid black line is the average curve (green area again marks the regime of physiological dephasing rates). In similarity to the cases of FMO and PC-645, one can see that indeed there is ENAQT, i.e. an increase in the exciton current, and that it is very small, $\sim 0.05$ percent. Similar results are obtained if multiple exciton injection and extraction are considered or if the exitations are of ring eigenstates (see SM, sections IV and V).   \\
\begin{figure}[H]
\centering
\includegraphics[width=1\linewidth]{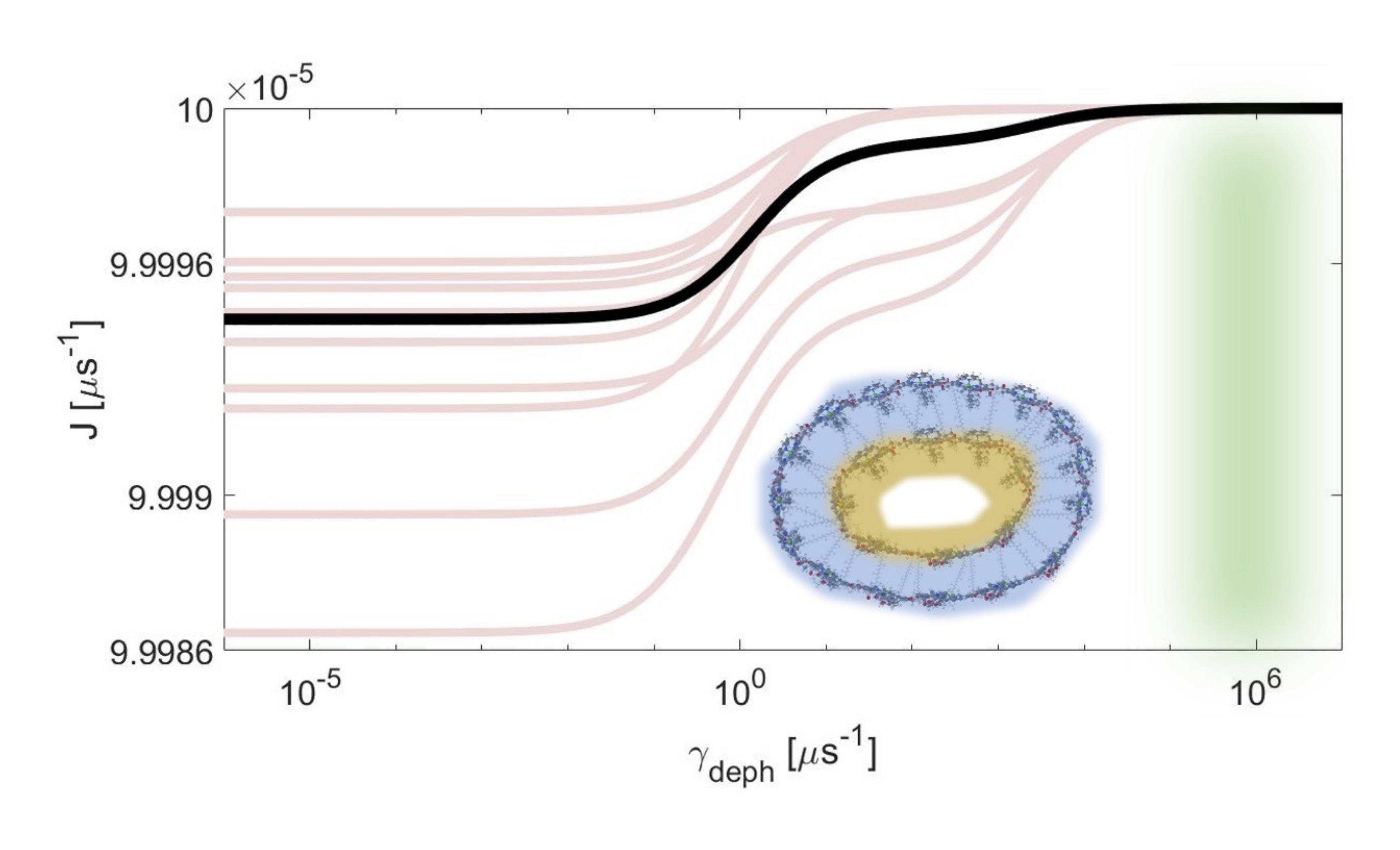}
\caption{Density configuration (i.e. exciton occupation at different sites) of the FMO complex for three different regimes: quantum limit (blue line, $\gamma_{deph}=10^{-4}\mu $s$^{-1}$), biological condition (yellow line,$\gamma_{deph}=10^6 \mu $s$^{-1}$) and quantum limit (green line, $\gamma_{deph}=10^{12} \mu $s$^{-1}$). The transition from the quantum regime towards the classical regime is accompanied by a shift in the density configuration, from a wave-function-determined configuration to a uniform gradient between the source and sink, with a uniform configuration in between \cite{zerah2018universal}. To see this more clearly, figures (b)-(c) present the schematic structure of FMO, where each sphere represents a BChl site, and the color brightness reflects its relative density. }
\label{fig:LHII_cur}
\end{figure}
\section*{ Discussion: Current,  coherence and classicality}
Figs 2 and 4 establish that there is no significant increase in the exciton current (and hence the efficiency) when comparing the fully quantum case (zero dephasing rate) and the physiological realistic dephasing rates ($10^6-10^8 \mu$s$^{-1}$). \\
Before discussing the relation between our results and the question in the title, we wish to elaborate further on the notion of "quantum" vs "classical". What makes the system "classical"? As pointed above, one definition would be the onset of a population gradient, i.e. classical diffusion and Fick's law \cite{zerah2018universal}, but this is an operational definition, which ignores the presence of coherences {(off diagonal terms of the density matrix)}\cite{yang_steady_state_2020}. Formally, coherence  is  necessary  for  observing  current,  because  the  current is proportional to (the imaginary part of) the off-diagonal elements of the density matrix, e.g., coherences. Thus, coherences are present also in the classical regime. \\
What really defines a "classical" system is not the lack of any coherence, but rather the fact that in a classical system, the coherences can be fully determined from the populations, without any additional information required. This implies no "long range" coherence, in the sense that sites which are not connected via hopping matrix elements in the Hamiltonian will have no coherence between them (i.e. no off-diagonal elements connecting them in the density matrix).  This distinction between quantum and classical dynamics was quantified in Ref.~\cite{wu_efficient_2012}, where the authors compared the local currents as derived from the off-diagonal density-matrix elements (i.e. quantum flux), $J^{Q}_{ij}=-2t_{ij} \Im (\rho_{ij})$ and from the diagonal elements (classical flux) $J^{C}_{ij}=\kappa_{ij}(\rho_{ii}-\rho_{jj})$ (where $t_{ij}$ is the hopping matrix element between sites $i$ and $j$, $\rho$ is the density matrix and $\kappa_{ij}$ is the classical exciton hopping rate between sites $i$ and $j$, see SM for further details). In a "classical" regime, the two currents would be the same, implying that quantum coherences  carry no additional information over the classical dynamics. This is what was found in Ref.~\cite{wu_efficient_2012} for the FMO complex. \\
%Since the term "coherence" is rather elusive and referred by many different definitions (because the value of the off diagonal elements is basis-dependent), we suggest the term "quantum correlations" to explain the situation where $J_{ij}^{Q}\ne J_{ij}^{C}$. One can then simply say that quantum correlations enhance efficiency if the total currents evaluated using $J^{Q}$ are higher than those evaluated using $J^{C}$.  }\\
Going back to our results, as can be clearly seen, there is a substantial drop in current when going towards very high dephasing rates. This occurs because in the classical regime, the system becomes diffusive, with a diffusion coefficient that is proportional to $\sim \bar{t}^2/\gamma_{deph}$ ($\bar{t}$ is some typical hopping matrix element)\cite{dziarmaga2012non}. The reduction of current with increasing $\gamma_{deph}$ (but never to zero) is simply due to the decrease in the diffusion coefficient. \\
One could then argue that the  intermediate dephasing rates observed in natural systems hold a substantial advantage over much higher dephasing rates. Since the system is "fully classical" at such high rates, but has substantial quantum coherence in the ENAQT regime, one would then argue that in fact quantum effects are very important in determining the efficiency. Put differently, one could say that evolution drove the design of the light-harvesting system (in structural parameters such as geometry, orientation, etc.) away from the classical regime in order to increase its exciton transfer efficiency.\\
To counter this argument, we note that the drop in current occurs at unrealistically high dephasing rates, which means that the inherent system parameters would have to be tuned (by evolution) in such a way that pushes the system to the classical regime in physiological dephasing times. However, we find that the regime of ENAQT is extremely robust, and depends very weakly on the Hamiltonian parameters (this can also be seen by the similarity between the FMO and PC-645 systems), in line with existing literature \cite{Shabani2014, hayes2011robustness, baker2015robustness,maiuri2018coherent}.\\
%To show this, we used a particle-swarm optimization algorithm \cite{kennedy1995particle} to optimize an FMO-like structure that minimizes the onset of classicality (see SI for further details). All Hamiltonian parameters were allowed to change, while keeping external parameters (i.e. injection and extraction rates) intact. We find that the minimal value of dephasing rate can only be as low as $\sim 10^9 \mu$s$^{-1}$, still much larger than the physiological dephasing rate (See SI section **).
This can be understood by analyzing, for instance, the analytical expressions for currents and populations of linear uniform models \cite{zerah2018universal} (see  SI section II). What we find is that the ENAQT regime is confined to the regime of $\gamma_{inj} < \gamma_{deph}< \frac{2\bar{t}^2}{\gamma_{inj}}$, where $\bar{t}$ is some effective or average hopping matrix element (which can be determined from, e.g. the bandwidth of the quantum system), and assuming $\gamma_{inj}\ll\gamma_{ext}$. Since the injection and extraction rates are external parameters, changing the ENAQT regime would require substantial reduction in the hopping matrix elements, but this would reduce the ability of the system to transfer energy. Presumably, this is the reason all three complexes have similar ranges of hopping matrix elements.\\
This analysis implies that the ENAQT regime is really the "natural" regime at which these systems operate; faster dephasing would require dynamics which are much faster than those of the proteins surrounding the ETCs, while longer dephasing times would require lower temperatures or deeper isolation of the chromophores. Figs.1 and 3 demonstrate that, surprisingly, at the ENAQT regime the current reaches a maximum which is limited by the injection rate, and is essentially independent of any Hamiltonian parameters. \\
To show this directly, in Fig. 5 we show the current-dephasing rate curves of over 5000 random realizations of FMO-like networks. In this calculation, the diagonal elements ( which have an absolute value with insignificant contribution to current) and the injection and extraction positions are kept fixed, and the hopping matrix elements, which define the network structure, are distributed randomly in the regime of $\pm 200 $cm$^{-1}$. The currents at the quantum regime differ substantially, since they depend on the detailed wave-function structure of a given realization (some realizations have very weak coupling between the source and sink, resulting in small currents at the quantum regime). Similarly, at the classical regime there is a distribution of currents (since the onset of the classical regime is sensitive to the hopping elements, and hence changes between realizations). However, at the ENAQT regime, there is essentially {\sl no dependence on the Hamiltonian parameters}.  This means that no matter how the network is arranged, the current is the same. In different words, at the ENAQT regime, the value of the current is completely indifferent to the network structure. Therefore, an inevitable conclusion is that enhancing the current was {\sl not} an evolutionary driver to determine the network structure. \\
This is not unique to the FMO complex. In fact, we find similar results for the PC-645 and LH2 complexes (not shown). Moreover, in the ENAQT regime the system is robust against not only geometrical changes, but also to many other parameters. As a specific example (one out of many), we consider the excitation points of LH2. The LH2 complex is coupled not only to external excitations (via direct light absorption) but also to excitation from neighboring complexes. To mimic this effect, in the inset of Figure~4 we plot the current-dephasing rate curves of the LH2 complex, taking a random number and position of injection and extraction sites (between two and four injection and extraction sites). Clearly, neither the number nor position of the injection sites affects the current in the ENAQT regime. \\
\begin{figure}[h]
\includegraphics[width=1\linewidth]{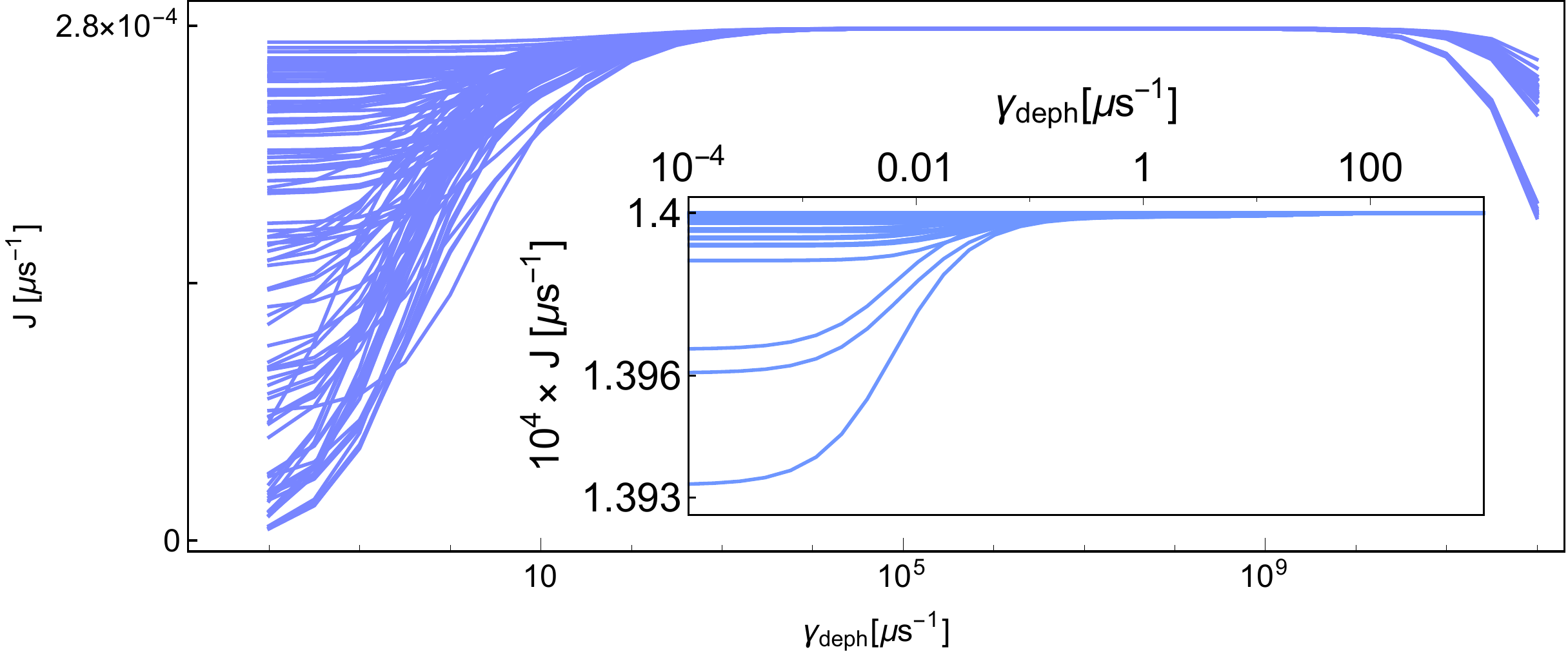}
\caption{Density configuration (i.e. exciton occupation at different sites) of the FMO complex for three different regimes: quantum limit (blue line, $\gamma_{deph}=10^{-4}\mu $s$^{-1}$), biological condition (yellow line,$\gamma_{deph}=10^6 \mu $s$^{-1}$) and quantum limit (green line, $\gamma_{deph}=10^{12} \mu $s$^{-1}$). The transition from the quantum regime towards the classical regime is accompanied by a shift in the density configuration, from a wave-function-determined configuration to a uniform gradient between the source and sink, with a uniform configuration in between \cite{zerah2018universal}. To see this more clearly, figures (b)-(c) present the schematic structure of FMO, where each sphere represents a BChl site, and the color brightness reflects its relative density.}
\label{fig:ModifiedFMO}
\end{figure}
The picture that emerges from the calculations presented here is as follows. The photosynthetic excitonic transfer networks, despite their structural differences, all operate in an environment with dephasing time $\tau_{deph}$ of a few hundred femtoseconds (up to 1ps), which puts them in the ENAQT regime. This regime is characterized by having an exciton current which is only slightly higher than the fully quantum regime, and reaches a maximum which is limited by the injection rate. The real advantage of being in the ENAQT regime is the surprising essential independence of the current on any particular structural parameters of the system. This leads to the conclusion that the structure of neither FMO, PC-645 or LH2 did not evolve in order to enhance efficiency (since it is essentially the same in all of them). \\
If anything, one could claim that evolution drove the physiological times to be what they are because in this regime the ETC network structure is irrelevant to efficiency, and hence can be used for a different function (stability, for instance  \cite{keren2018photosynthetic}). However, one must also consider the possibility that the physiological coherence time was also not part of the evolutionary driving, and developed by nothing more than a happy coincidence. Indeed, the physiological dephasing time seems to be a middle ground; faster coherence is typical to electronic systems but not to vibrational systems, and slower coherence times are unlikely in physiological environment. \\
Therefore, two central challenges remain for future studies. The first is still to understand what determines the origin of the dephasing time $\tau_{deph}$ observed in experiments (e.g., \cite{pachon2011physical,zhang2016origin}),  with the goal of understanding whether the observed values are somehow unique and could they be different. The second challenge is to look for other possible evolutionary advantages that the structures of the photosynthetic transfer complexes provide. Overcoming these challenges will push forward the understanding of the possible role of quantum effects in photosynthetic complexes. 
\section*{Acknowledgments}
E.Z.H Acknowledges support from the Ilse Katz Center center interdisciplinary fellowship. The authors are grateful to Michael Zwolak for valuable discussions. \textbf{Funders:} This work was supported by the Israel Science Fund grant No.~1360/17. \textbf{Data and materials availability:} All data needed to evaluate the conclusions in the paper are present in the paper and/or the Supplementary Materials. \textbf{Contribution:}
EZH and YD both conceived and performed the research, and wrote the paper. \textbf{Competing interests:} The authors declare no competing interests.

%Here you should list the contents of your Supplementary Materials -- below is an example. 
%You should include a list of Supplementary figures, Tables, and any references that appear only in the SM. 
%Note that the reference numbering continues from the main text to the SM.
% In the example below, Refs. 4-10 were cited only in the SM.     
\section*{Supplementary materials}
%Section 1: Detailed description of the computational procedure: Lindblad equation and ETC Hamiltonians.\\
%Section 2: Analytic description of ENAQT\\
%Section 3:ENAQT enhancement at the extreme parameter range\\
%Section 4: Multiple excitations and injections in LHII\\
%Section 5: Excitation in the energy basis\\
%Section 6: Detailed Hamiltonian\\
Fig.s1. Current  enhancement  as  a  function  of  injection and  extraction rates for FMO and pc-645\\
Fig.s2. Histogram of current enhancement from 10,000 different configurations of injection and extraction sites.\\
Fig.s3.  Current as a function of dephasing rate for the case for excitation in the energy basis.\\

% For your review copy (i.e., the file you initially send in for
% evaluation), you can use the {figure} environment and the
% \includegraphics command to stream your figures into the text, placing
% all figures at the end.  For the final, revised manuscript for
% acceptance and production, however, PostScript or other graphics
% should not be streamed into your compliled file.  Instead, set
% captions as simple paragraphs (with a \noindent tag), setting them
% off from the rest of the text with a \clearpage as shown  below, and
% submit figures as separate files according to the Art Department's
% instructions.

%\bibliography{MyBib}

%\bibliographystyle{ScienceAdvances.bst}

\clearpage

\noindent {\bf Fig. 1.} 
Mechanisms of efficiency-driven evolution and environment-assisted quantum transport (a) Schematic description of the evolutionary progress of photosynthetic complexes towards their current geometry, with efficiency being the evolutionary driving force. As evolution progresses, the structure of the photosynthetic complex evolves towards its current structure (the FMO complex in this example), while increasing efficiency. Is this indeed the evolutionary pathway of photosynthetic complexes, and if so is quantum coherence part of the efficiency enhancement is a central question in the field of quantum biology. 
(b) Schematic depiction of the population uniformization mechanism shown for a uniform chain of six sites (Blue lines depict the sites in the chain, Yellow arrows show the excitation of first site and extraction from fifth site). The density of the sites are described by blue bars for the quantum regime,  ENAQT regime and  classical regime, along with a schematic form for the current vs dephasing curves.  
 \\
\\
\noindent {\bf Fig. 2.} 
Calculated exciton current as a function of dephasing for the FMO (a) and pc-645 (b) complexes. The shaded Green area indicates the estimated range of physiological dephasing rates. Insets show a schematic description of the exciton complexes (the full Hamiltonians used are provided in the SM section VI).\\
\\
\noindent {\bf Fig. 3.} 
Density configuration (i.e. exciton occupation at different sites) of the FMO complex for three different regimes: quantum limit (blue line, $\gamma_{deph}=10^{-4}\mu $s$^{-1}$), biological condition (yellow line,$\gamma_{deph}=10^6 \mu $s$^{-1}$) and quantum limit (green line, $\gamma_{deph}=10^{12} \mu $s$^{-1}$). The transition from the quantum regime towards the classical regime is accompanied by a shift in the density configuration, from a wave-function-determined configuration to a uniform gradient between the source and sink, with a uniform configuration in between \cite{zerah2018universal}. To see this more clearly, figures (b)-(c) present the schematic structure of FMO, where each sphere represents a BChl site, and the color brightness reflects its relative density.
\\
\\
\noindent {\bf Fig. 4.}
Average LH2 exciton current as a function of dephasing rate (black line),  calculated for $\approx 900$ possible paths. Pink curves show the current of arbitrary chosen realizations (i.e entry and exit sites) in LH2. Shaded green area marks the natural dephasing rate. Inset:  Schematic description of LH2 transfer network (the full Hamiltonian used is provided in the SM section VI).\\
\\
\noindent {\bf fig. 5.}
Current vs dephasing rate for 5000 realizations of FMO-like networks. Energies were kept fixed while hopping matrix elements were picked from a range of $\pm200$cm$^{-1}$. ENAQT is obtained for almost the same range for all realisations, indicating the independence of efficiency in the ENAQT regime (and the regime itself) on the structure of the system.

\end{document}